\documentclass[showpacs,twocolumn,prl]{revtex4-1}
\usepackage{amssymb,amsmath,amsthm,mathtools,color,array}
\newcolumntype{P}[1]{>{\centering\arraybackslash}p{#1}}

\begin{document}
\title{Wetting Transitions Displayed by Persistent Active Particles}

\author{N\'estor Sep\'ulveda}
\author{Rodrigo Soto}
\affiliation{Departamento de F\'{\i}sica, Facultad de Ciencias F\'{\i}sicas y Matem\'aticas, Universidad de Chile, Avenida Blanco Encalada 2008, Santiago, Chile}

\date{\today}

\begin{abstract}
A lattice model for active matter is studied numerically, showing that it  displays wettings transitions between three distinctive phases when in contact with an impenetrable wall. 
The particles in the model move persistently,  tumbling with a small rate $\alpha$, and interact via exclusion volume only. When increasing the tumbling rates $\alpha$, the system transits from total wetting to partial wetting and unwetting phases. In the first phase, a wetting film covers the wall, with increasing heights when $\alpha$ is reduced. The second phase is characterized by wetting droplets on the wall with a periodic spacing between them. Finally, the wall dries with few particles in contact with it.
These phases present nonequilibrium transitions. The first transition, from partial to total wetting, is continuous and the fraction of dry sites vanishes continuously when decreasing  the tumbling rate $\alpha$. For the second transition, from partial wetting to dry, the mean droplet distance diverges logarithmically when approaching the critical tumbling rate, with saturation due to finite-size effects.
\end{abstract}
\maketitle

{\em Introduction.}
Active matter, composed of elements  able to transform stored energy in a reservoir into motion at the individual scale, has gained attention as a model for nonequilibrium systems where the driving is internal. It is a conceptual framework that aims to describe the collective and individual dynamics of living matter (microtubules, bacteria and other microorganisms, fish, birds, or even larger animals) as well as nonliving objects (active colloids, vibrated grains, etc.).
One of the properties that characterize active matter is that particles present self-propulsion and persistence; this gives rise to several phenomena, such as phase separation, large diffusivities, giant density fluctuations, etc\ (see Refs.~\cite{Vicsek2012, Marchetti2013,bialke2015} for recent reviews and references therein). Different models have been introduced to account for persistence, where active Brownian and run-and-tumble particles are the most used for systems lacking of inertia, both in and off lattice~\cite{Berg,Romanczuk2012,solon2015}. Depending on the model, persistence is characterized by either the correlation length, the rotational diffusivity, or the tumbling rate.

Self-propelled particles display important interactions with solid surfaces, which normally manifest in particle accumulation at the walls. At low Reynolds numbers, there is a hydrodynamic attraction for pusher swimmers, which leads to  accumulation~\cite{Berke2008,Elgeti2009,Dunstan2012,Elgeti2013,Molaei2014,Mathijssen2016}. Also, the geometrical alignment with surfaces that results from  collisions implies that particles remain in their vicinity for long times~\cite{Li2009,Elgeti2015,Ezhilan}. Finally, concave walls, by their shape, generate longer residence times near surfaces~\cite{Vladescu2014,Cheng2015,Wysocki2015}, an effect that is also present for {\em E.coli} at convex walls provided that the curvature radius is large enough~\cite{Sipos2015,Wysocki2015}. In all these cases, the accumulation at surfaces is a result of individual interactions.  Collective effects have also been considered in  Refs.~\cite{Wensink2008,Costanzo2012,Figeroa2015}.
In this Letter, we are interested in the collective interaction of self-propelled particles with solid surfaces. In particular, we are interested in the development of wetting transitions.

Molecular systems at equilibrium, when interacting with solid surfaces, can present several wetting states, depending on the temperature and surface tension~\cite{deGennes}. In particular, a transition between phases of partial wetting, with finite contact angle, and total wetting, where the contact angle vanishes strictly, takes place close to the bulk critical point~\cite{Cahn,Moldover}. This equilibrium wetting transition presents universal critical properties~\cite{Nakanishi,deGennes}.

To elucidate whether a collection of active particles develops wetting transitions similar to those at equilibrium, we study the simple persistent excluding particles (PEP) model~\cite{RG2014,SS2016}, where particles interact by excluded volume only.
In the PEP model, particles move in a regular lattice at discrete time steps with directions that change randomly (tumble) at a small rate and excluded volume is achieved by imposing a maximum occupation per site.  Persistence is characterized by the tumbling rate $\alpha$, where the correlation length scales as $\alpha^{-1}$. Depending on the tumble rate and the maximum occupancy per site,  particles can either aggregate in few big clusters after a coarsening process,  develop many  small clusters, or form an homogeneous gas, with nonequilibrium transitions between these phases~\cite{SS2016}.

Here, we show that in the presence of a impenetrable wall, three phases can develop. At low tumbling rates, the particles can  present total wetting, where a thick film of particles forms. Increasing the tumbling rate, a transition to partial wetting takes place. Here, droplets dispose almost periodically on the wall, increasing their distance when increasing the tumbling rate. Finally, at higher tumbling rates, the wall dewets of particles. These two transitions present critical behavior, and critical exponents for appropriate order parameters are obtained. 

\begin{figure*}
\includegraphics[width=2.\columnwidth]{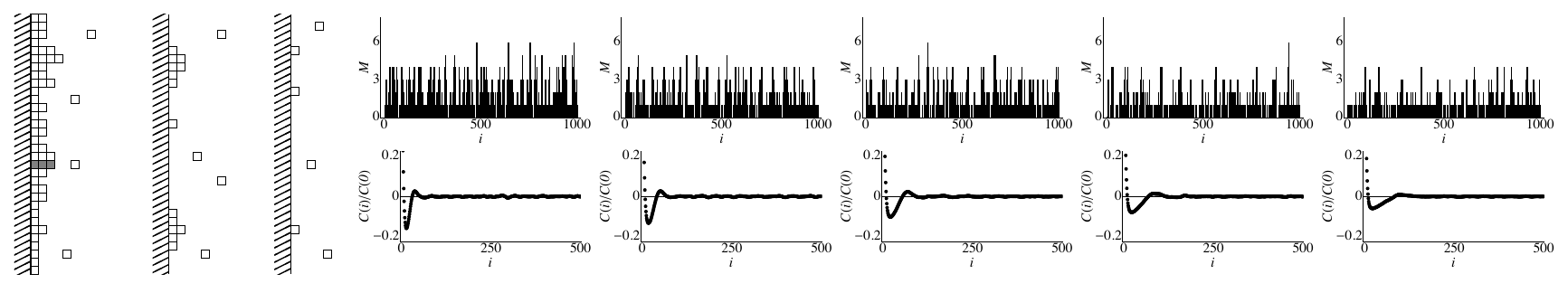}
\caption{(Left) Schematic representation of the three phases (total wetting, with fraction of empty sites $Z=0$; partial wetting, with $0<Z<1$; and dewetted of particles, with $Z\approx 0$). In gray, a film column with $\Delta=3$. (Right, top) Instantaneous values of the mass profile $M(i)$  for $n_\text{max}=3$ and increasing values of 
$\alpha=0.0067, 0.0092, 0.0115, 0.0137, 0.0160$, for instants selected when the system presents the partial wetting phase. 
Similar figures are obtained for $n_\text{max}=4$. 
(Right, bottom)
Mass correlation function $C(i)/C(0)$ for the same parameters.  Recall that periodic boundary conditions are used on $y$, hence limiting $i$ to $L_y/2$ for the evaluation of $C$. }
\label{fig.correlation}
\end{figure*}

{\em Model.}
We consider a two-dimensional realization of the PEP model. A total of $N$ particles move on a regular lattice composed of $L_x\times L_y$ sites. Each particle has a state variable $\mathbf{s} =\{\pm \hat{\mathbf{x}}, \pm \hat{\mathbf{y}}\}$, which indicates the direction to which it points. Time evolves at discrete steps. At each step, particles attempt to jump one site in the direction pointed by $\mathbf{s}$. 
If the occupation of the destination site is smaller than the maximum occupancy per site $n_\text{max}$, the jump takes place; otherwise, the particle remains at the original position.  $n_\text{max}$ models particle overlaps: if $n_\text{max}=1$  the interaction between particles is steric, while larger values allow for increasing degrees of overlap. This overlap can account for the deformation  of microorganisms or their organization in quasi-two-dimensional layers, which allows larger occupancies in the two-dimensional  projected plane.
The position update of the particles is asynchronous to avoid two particles attempting to jump to the same site simultaneously. That is, at each time step, the $N$ particles are sorted randomly and, sequentially, each particle jumps to the site pointed by ${\mathbf s}$ depending on its occupation.  
Finally, at the end of the update phase, tumbles are performed: for each particle, with probability $\alpha$~\cite{commentalpha}, the director ${\mathbf s}$ is redrawn at random from the four possibilities, independently of the original value. 
This model (model 1) is the two-dimensional version of the one presented in Refs.~\cite{RG2014,SS2016}. 
In analogy to hard-core systems, particles only interact via excluded volume effects and there are no energetic penalizations, explicit partial motility reduction as in Motility-Induced Phase Separation~\cite{cates2015}, or explicit alignment as in Ref.~\cite{peruani2011}. 
 
To study the wetting phenomenon, the system is bounded by walls in the $x$ direction, while being periodic in $y$. Particles pointing to a wall cannot jump to it and remain immobile  until tumble events take place. For low tumbling rates, there is a high probability that new particles arrive, blocking the first ones. The wall therefore acts as a nucleation site for clusters and, naturally, wets by particles. The wetting film will increase by deceasing $\alpha$, as has been observed under dilute conditions \cite{Elgeti2015,Ezhilan}. Here, we investigate if this tendency is smooth or, rather, if transitions take place.

The control parameters are the particle density $\phi=N/(L_x L_y)$, the tumble rate $\alpha$, and the maximum occupation number $n_\text{max}$. Time is measured in time steps and lengths in lattice sizes; therefore particles have velocities $0,\,\pm\hat{\mathbf{x}},\, \pm\hat{\mathbf{y}}$. The model has the intrinsic time scale corresponding to the time it takes a particle to cross one site $t_\text{cross}=1$ and also the mean flight time, which depends on density. Tumbling introduces a new  time scale $t_\text{tumble}=1/\alpha$, which must be compared with the previous two, rendering the phase diagram highly nontrivial.

{\em Wetting transitions.}
Simulations for models 1, 2, and 3 are performed in  the  range of the parameters  $n_\text{max}=1,2,3,4$ and $0.025\times10^{-2}\leq\alpha\leq 3.02\times10^{-2}$,  
while the number of sites and global density has been fixed to $L_x=6000$, $L_y=1000$, and $\phi=0.01$. For model 4, $0.025\times10^{-2}\leq\alpha\leq 6.23\times10^{-2}$, $L_x=3000$, $L_y=500$, and $\phi=0.04$. Models 2, 3, and 4 are defined below.
We have verified by simulations that in absence of walls, the system remains in the gas phase for these parameters. This choice for the system size has the following rationale. First, $L_y$ must be large enough to detect any spatial structuration of the wetting film that will be formed on the walls. Second, if $L_x$ is small, particles that evaporate after a tumble from a wall could arrive to the other wall ballistically, creating artificial correlations between the two walls. To ensure that evaporated particles reach the diffusive regime, one must take $L_x\gg\alpha^{-1}$.  Initially, particles are placed randomly, with random orientations. 
Finally, we consider total simulation times equal to $T=2\times 10^6$ time steps, which are long enough to reach a steady state and to achieve good statistical sampling. 

The wetting films are characterized as follows. For each vertical position $i$, the film column is defined as the contiguous nonempty sides beside the wall (see Fig.~\ref{fig.correlation}). From this, we measure the instantaneous thickness profile $\Delta(i)$ and the total number of particles $M(i)$ in the column.
By observing the instantaneous profiles, three distinct phases are recognized [as a reference, Fig.~\ref{fig.correlation} displays the instantaneous mass profiles $M(i)$ for $n_\text{max}=3$ and various values of $\alpha$].
For low tumbling rates, the thickness is uniformly greater than 1, corresponding to a phase of total wetting. Increasing $\alpha$, the average values of $M$ and $\Delta$ decrease. For intermediate $\alpha$, droplets  appear (roughly arranged periodically) where regions of vanishing and finite values of $M$ alternate, corresponding to partial wetting. Note that,  contrary to equilibrium wetting transitions, the droplets are 1 or 2 sites  thick and no visible spatial structuration is observed in $\Delta$. Therefore, it is not possible to define and compute a contact angle, nor it is possible to describe them using classical growth models as the Kardar, Parisi, and Zhang equation~\cite{KPZ}. Finally, for larger values of $\alpha$, the thickness is small throughout the wall, which is interpreted as a dewetted  phase. 
The emergence of droplets is a consequence of the particles' motility and persistence. When a particle hits the wall, it remains there until a tumble takes place. If, after a tumble, it moves parallel to the wall, it will block when hits another particle, acting then as a nucleation site for droplets. The balance between evaporation and the incoming flux of particles moving along the wall stabilizes the size of droplets. 
Though the bulk density is small, the density in contact with the wall is large, reaching the maximum value $n_\text{max}$. Varying $\phi$ changes the transition tumbling rates, but the phenomenology and the morphology of the wetting film and the droplets remain, as is shown for example in  model 4, where $\phi=0.04$.

\begin{figure*}
\includegraphics[width=2.\columnwidth]{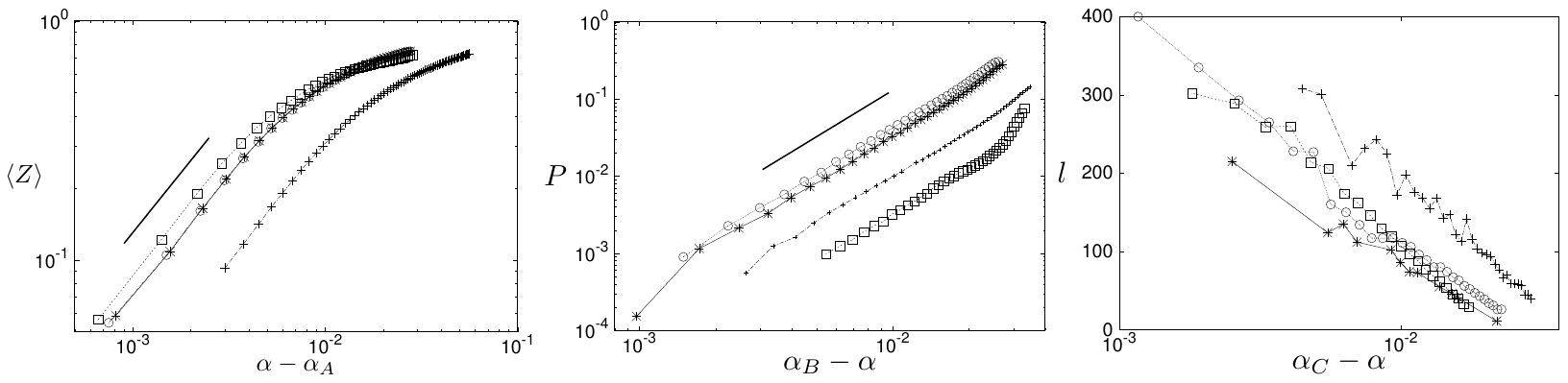}
\caption{
Critical behavior of the three order parameters for the lattice models 1 (stars), 2 (squares), 3 (circles), and 4 (crosses). 
(Left) $\langle Z\rangle$ above the total to partial wetting transition for $n_\text{max}=2$. (Middle, right) $P-p_0$ and $l$ below the partial wetting to dewetting transition for $n_\text{max}=3$.  The simulation results, with symbols, are joined by lines to assist in visualization. The solid lines in the left and middle panels present  power laws with exponents $1$ and $2$, respectively. The order parameters for larger values of $n_\text{max}$ present similar critical behavior (not shown).}
\label{fig.critical}
\end{figure*}

To analyze the first transition, from total to partial wetting, we compute the instantaneous fraction of sites with vanishing thickness, $Z=|\{i : \Delta_i=0\}|/L_y$.  Its temporal average $\langle Z\rangle$ is an order parameter that for $n_\text{max}\geq 2$  vanishes when total wetting is achieved and presents a continuous transition [see Fig.~\ref{fig.critical}(left)].
No transition takes place for $n_\text{max}=1$. Close to the critical point it vanishes as 
\begin{align}
\langle Z\rangle &= Z_0 (\alpha-\alpha_A)^{1.0\pm 0.1}, & \alpha&\geq \alpha_A,  \label{eq.fitZ}
\end{align}
where the critical values are indicate in Table \ref{tablaVC}, while $Z_0$ are nonuniversal prefactors. The exponent $1.0$ is a best fit to power laws, using the protocol described in Ref.~\cite{Castillo}, which allows for the extraction of the critical tumbling rates.

\begin{figure*}[htb]
\begin{center}
\includegraphics[width=2.\columnwidth]{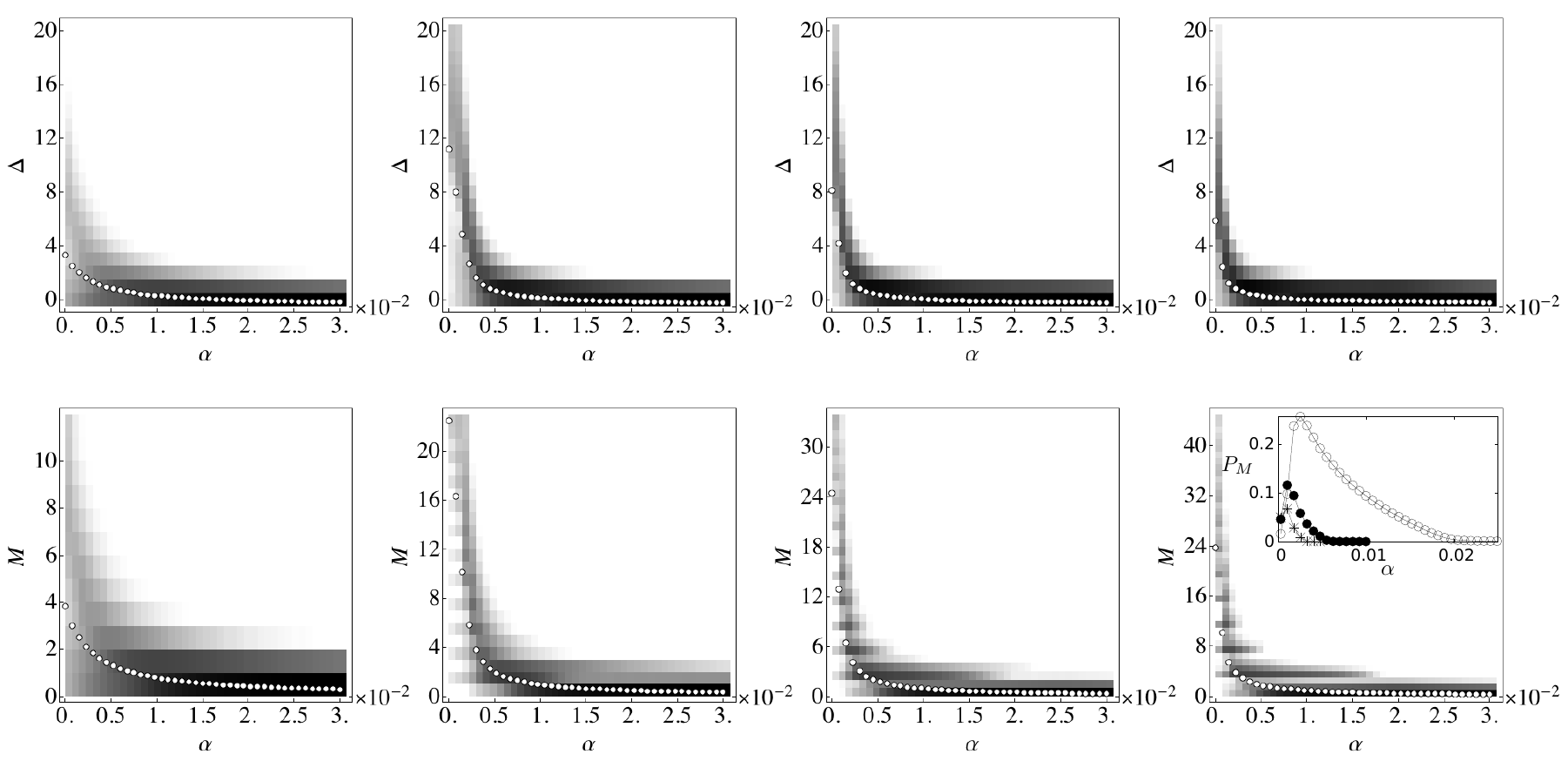}
\end{center}
\caption{Probabilities of obtaining different values of  the order parameters $\Delta$ (upper row) and $M$ (lower row), for different values of $\alpha$, and for $n_\text{max}=1,2,3,4$ (ordered from left to right). Darker levels of gray indicate larger probabilities. Dots indicate the average value of the order parameters for each value of $\alpha$. Inset: Probabilities for $M=n_\text{max}$ (open circles), $M=2n_\text{max}$ (full circles), and $M=3n_\text{max}$ (stars).}
\label{histogramas}
\end{figure*}

A detailed analysis of the instantaneous values of $\langle \Delta\rangle$ and $\langle M\rangle$ shows that the second transition, from partial wetting to dewetting, presents bistability between these two phases, which alternate in the course of time.
Therefore, for each value of $n_\text{max}$ and $\alpha$, we  measure the order parameters every 1000 time steps and build the histograms of the different values they can take. As $M$ and $\Delta$ take integer values, no binning is necessary. These probability distributions are shown in Fig.~\ref{histogramas}. For $n_\text{max}=3,4$, the distributions of $M$ show that  the system presents multistability, with higher probabilities having an integer number of saturated layers ($M$ being a multiple number of $n_\text{max}$). An analysis of the probability distributions shows that, indeed, there are different transitions, when the probabilities of having a certain number of layers vanish (see Fig.~\ref{histogramas}, inset). 
For $n_\text{max}=1$, no transition takes place, but there is instead a continuous decrease of the different probabilities when increasing $\alpha$, and the case for $n_\text{max}=2$ is less clear. This behavior is similar to the clustering transitions in one dimension, where clear transitions are also observed for $n_\text{max}\geq 3$; for $n_\text{max}=1$ no transition takes place and $n_\text{max}=2$ corresponds to a crossover \cite{SS2016}.

We analyze in detail the case when the last layer ($M=n_\text{max}$) disappears, as it corresponds to the transition to dewetting. Figure~\ref{fig.critical}(middle) shows that $P\equiv P(M=n_\text{max})$ vanishes at a critical tumbling rate $\alpha_B$, following a power law
\begin{align}
P &= p_0+p_1(\alpha_B-\alpha)^{2.0\pm0.2}, & \alpha&<\alpha_B, \label{eq.fitP}
\end{align}
where the exponent 2 is a best fit, $p_0$ is a small offset probably due to finite-size effects, $p_1$ is a nonuniversal value, and the critical values are given in Table~\ref{tablaVC}.

\begin{table}
\caption{ Critical behavior obtained for the different order parameters (first column) and models (second column). Best-fit values for the universal exponent (third column) and nonuniversal critical values of $\alpha$ for $n_\text{max}=2$ in the case of $\langle Z\rangle$ and for $n_\text{max}=3$ in the case of $P$ and $l$. The values in the last column are multiplied by $10^{3}$ (mean value and error).}
\label{tablaVC}
%\begin{center}
\begin{tabular}{P{0.5cm} P{.5cm} P{1.25cm} P{2.cm} }  
\hline
$\langle Z\rangle$   &  1 & $1.0\pm 0.1$ & $2.42\pm0.05$ \\
          &  2 &                          & $1.83\pm 0.05$        \\
          &  3 &                          & $2.49\pm 0.05$        \\
          &  4 &                          & $5.46\pm 0.05$        \\
\hline
$P$   &  1 & $2.0\pm 0.2$    & $30\pm1$\\
          &  2 &                          & $36\pm1$ \\
          &  3 &                          & $29\pm1$ \\
          &  4 &                          & $43\pm1$\\
\hline
$l$&  1 & $\ln$                     & $24\pm2$ \\
          &  2 &                         & $21\pm2$ \\
          &  3 &                         & $28\pm2$ \\
          &  4 &                         & $37\pm2$ \\
\hline
\end{tabular}
%\end{center}
\end{table}

To analyze the spatial structuration in droplets in the partial wetting to dewetting transition, we consider  the mass spatial correlations $C(i) \equiv \langle M(i+j) M(j)\rangle -\langle M\rangle^2$. The average is done in the steady state over $j$ and time. 
Figure \ref{fig.correlation} displays the correlation function together with instantaneous values of the mass profile $M(i)$, for   $n_\text{max}=3$ and various values of $\alpha$. 
It is evident that a periodic structure is present for low values of $\alpha$. These correspond to wetting droplets condensed on the walls. A similar behavior is observed for $n_\text{max}=4$ (not shown). When increasing $\alpha$, the associated wavelength increases until it saturates with the system size $L_y$. The position $l$  of the first maximum of the correlation functions, which measures the typical distance between droplets, is analyzed as a function of $\alpha$ for fixed $n_\text{max}$, showing the existence of a critical point $\alpha_C$. For $\alpha<\alpha_C$, the critical law
\begin{align}
l &= l_0- l_1 \ln(\alpha_C-\alpha)   \label{eq.fitl}
\end{align}
is obtained, as shown in Fig.~\ref{fig.critical}(right). Here $l_0$ and $l_1$  are nonuniversal prefactors. The critical values in Table~\ref{tablaVC} are compatible with the corresponding values of $\alpha_B$, extracted from $P$, which characterize the same transition. The difference between both critical tumbling rates is due to finite-size effects and it is a usual effect that  appears when analyzing critical exponents.

{\em Universality.} 
To study the universality of the previous results, we consider three variations to model 1. The first variation (model 2) consists of jumps taking place probabilistically as in Ref.~\cite{SS2016}, where the jump probability depends on the destination site occupation and $n_\text{max}$. In the second variation (model 3), when a tumble takes place, the new director $\mathbf{s}$ is sorted at random between the two directions perpendicular to the pretumble director, introducing a short time memory in the dynamics. Finally, the third variation (model 4) consists of using a triangular lattice, with a major axis parallel to the wall.
In these three cases, we perform the same analysis on the order parameters; we find different values for the nonuniversal coefficients and critical tumble rates, but the exponents are equal to those of model 1 (see Fig.~\ref{fig.critical} and Table~\ref{tablaVC}). 

{\em Conclusions and discussion.} 
We have numerically studied the steady states of four models for active matter in presence of an impenetrable wall. Wetting phases are found, analogous to those present in equilibrium molecular fluids. Namely, the system can present total wetting, partial wetting, and dewetting phases. The control parameter is the tumbling rate, which plays an analogous role to temperature, allowing particles to evaporate from the wall. Between these phases, nonequilibrium phase transitions with power-law scaling for the order parameters are obtained.  We verified that  the critical exponents are universal, within the range of lattice models. 
A comparison with other models for active matter, for example in off-lattice models, will indicate whether the obtained critical laws and the structure of the phase diagram are universal.
The possible relation between the clustering transitions in a lower-dimensional space (the wall) reported in Ref.~\cite{SS2016} and the wetting transitions deserves further study. However, there is a substantial difference, in that the latter presents two consecutive transitions absent in the case of clustering.
Finally, different wetting states can have  important effects on the early stages of biofilm formation or other collective aggregation at surfaces, as these states differ appreciably in the particle disposition at the surface, creating periodic droplets or a uniform film.

%%%%%%%%%%%%
{\em Acknowledgments.} 
This research was supported by Fondecyt Grants No.\ 1151029 (N.S.) and No. 1140778 (R.S.).

%%%%%%%%%%%%


\begin{thebibliography}{99}
\bibitem{commentalpha}
Strictly speaking, for a tumbling rate $\alpha$,  the tumbling probability at each time step is $(1-e^{-\alpha})$. For small values, as we consider in this Letter, it is well approximated by $\alpha$.
\end{thebibliography}
\end{document}